\documentstyle[11pt,newpasp,twoside]{article}
\input{psfig}
\def\HI{H{\,\small I}}
\newcommand{\ltsima} {$\; \buildrel < \over \sim \;$}
\newcommand{\gtsima} {$\; \buildrel > \over \sim \;$}
\newcommand{\lta} {\lower.5ex\hbox{\ltsima}}
\newcommand{\gta} {\lower.5ex\hbox{\gtsima}}

\newcommand{\kms}{$\,$km$\,$s$^{-1}$}
\def\edcomment#1{\iffalse\marginpar{\raggedright\sl#1\/}\else\relax\fi}
\reversemarginpar

\begin{document}
\title{Gas outflows in radio galaxies}
 \author{R. Morganti, T. Oosterloo}
\affil{Netherlands Foundation for Research in Astronomy, Postbus 2,
NL-7990 AA, Dwingeloo, NL}
\author{B.H.C. Emonts}
\affil{Kapteyn Astronomical Institute, RuG, Landleven 12, 9747 AD,
Groningen, NL}
\author{C.N.  Tadhunter, J. Holt}
\affil{Dept.\ Physics \& Astronomy,
University of Sheffield, S7 3RH, UK}

\begin{abstract}

We present a summary of our recent results on gas outflows in radio
galaxies.  Fast outflows (up to 2000 \kms) have been detected both in
ionized and neutral gas. The latter is particularly surprising as it
shows that, despite the extremely energetic phenomena occurring near
an AGN, some of the outflowing gas remains, or becomes again,
neutral. These results are giving new and important insights on the
physical conditions of the gaseous medium around an AGN.

\end{abstract}

\section{Introduction}

Fast nuclear gas outflows appear to be a relatively common phenomena
in active galactic nuclei (AGNs).  They have been detected - using
optical, UV and X-ray observations - in Seyfert galaxies (see for some
examples Crenshaw et al.\ 2000; Aoki et al.\ 1996; Capetti et al. 1999
and refs therein), quasars (see e.g.\ Turnshek 1986, Krongold et al.\
2003 and refs therein), but also in starburst galaxies (Veilleux et
al.\ 2002, 2003).  These gas outflows can have different origins, such as
being jet-driven (if the object is radio loud) or related to starburst
and AGN winds.
 
Why should we expect gas outflows also in radio galaxies? In these
objects all the potential drivers for powerful outflows (a highly
luminous ionising AGN, relativistic jets and, often, a starburst) are
present to varying degrees.  In particular, the activity-inducing
processes (e.g.\ merger, interaction) are believed to involve the
injection of substantial amounts of gas/dust into the central nuclear
regions. Thus, we can expect that in the first phase (or in a
re-started phase) of activity, strong interactions with the relatively
dense medium will occur. Indeed, gas outflows are now increasingly
detected in radio galaxies and here we present a summary of the
results  we have obtained so far in our study of both the ionized,
as well as the neutral, components of such outflows.

\begin{figure*}
\centerline{
\psfig{figure=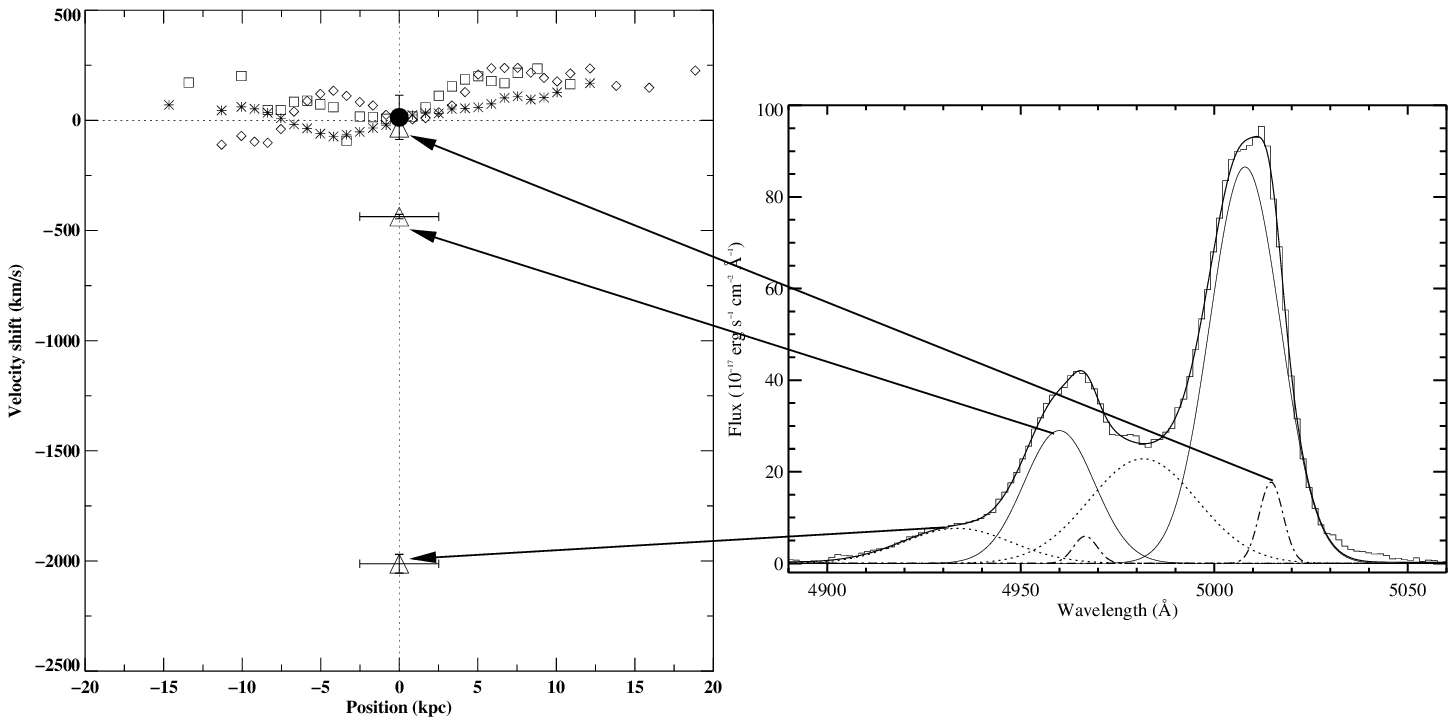,angle=0,width=8cm}
\psfig{figure=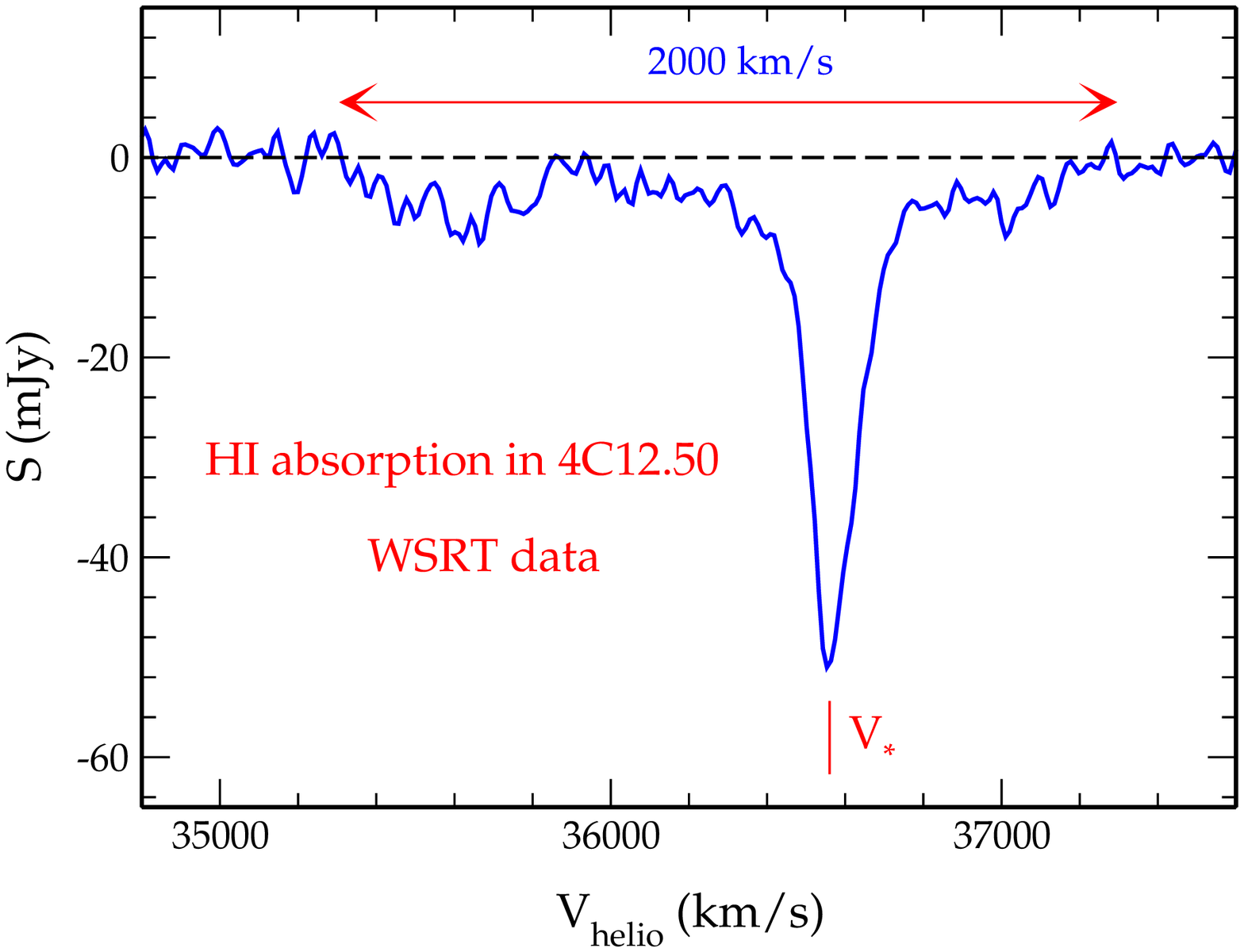,angle=0,width=6cm}}
\caption{{\sl Left} Radial velocity profile of the extended gaseous
halo of 4C12.50. Key: $\ast$, $\Diamond$ and $\Box$: highly extended
{[O II]} emission (narrowest component); $\bullet$: \HI\ 21-cm
absorption; $\triangle$: 3 components of {[O III]} in
the nucleus.  The [OIII] profile with the components from the fit is
also shown.  {\sl Right} The \HI\ absorption profile detected in
4C~12.50 from the WSRT observations.  The spectra are plotted in flux density
(mJy) against optical heliocentric velocity in \kms.  }
\end{figure*}

\section{Outflows of ionized gas}

The first clear example of an outflow of ionized gas that we have found is in
the southern {\sl starburst} radio galaxy PKS~1549--79 (i.e.\ a radio galaxy
that spectroscopically shows a young stellar population component in addition
to the old stellar component typical of elliptical galaxies).  In
PKS~1549--79, different redshift systems associated with both the low- and
high-ionisation emission lines were found (Tadhunter et al.\ 2001).  The high
ionisation lines are broader, with FWHM$> 1000$ \kms\ while the low-ionisation
lines are narrower. The \HI\ absorption detected in this object (Morganti et
al.\ 2001) has the same redshift as the low ionisation gas, while the high
ionisation gas appears blueshifted with respect to it. The \HI\ absorption and
the low ionisation gas are associated with a cocoon of material surrounding
the tiny ($\sim 200$ pc in size) radio source.  On the other hand, the high
ionisation (and blueshifted) gas is believed to be associated with an outflow
from a region close to the radio jet and disturbed by it. Taken together, all
these results are consistent with the idea that PKS~1549--79 is a young radio
source in which the cocoon of debris left over from the event that triggered
the activity has not yet been swept aside by circumnuclear outflows.  More
recently, Bellamy et al.\ (2003) have detected broad Pa$\alpha$ in
PKS~1549--79 confirming that this object indeed contains a quasar nucleus that
is moderately extinguished, despite evidence that its radio jet points close
to our line-of-sight.

\begin{figure*}
\centerline{\psfig{figure=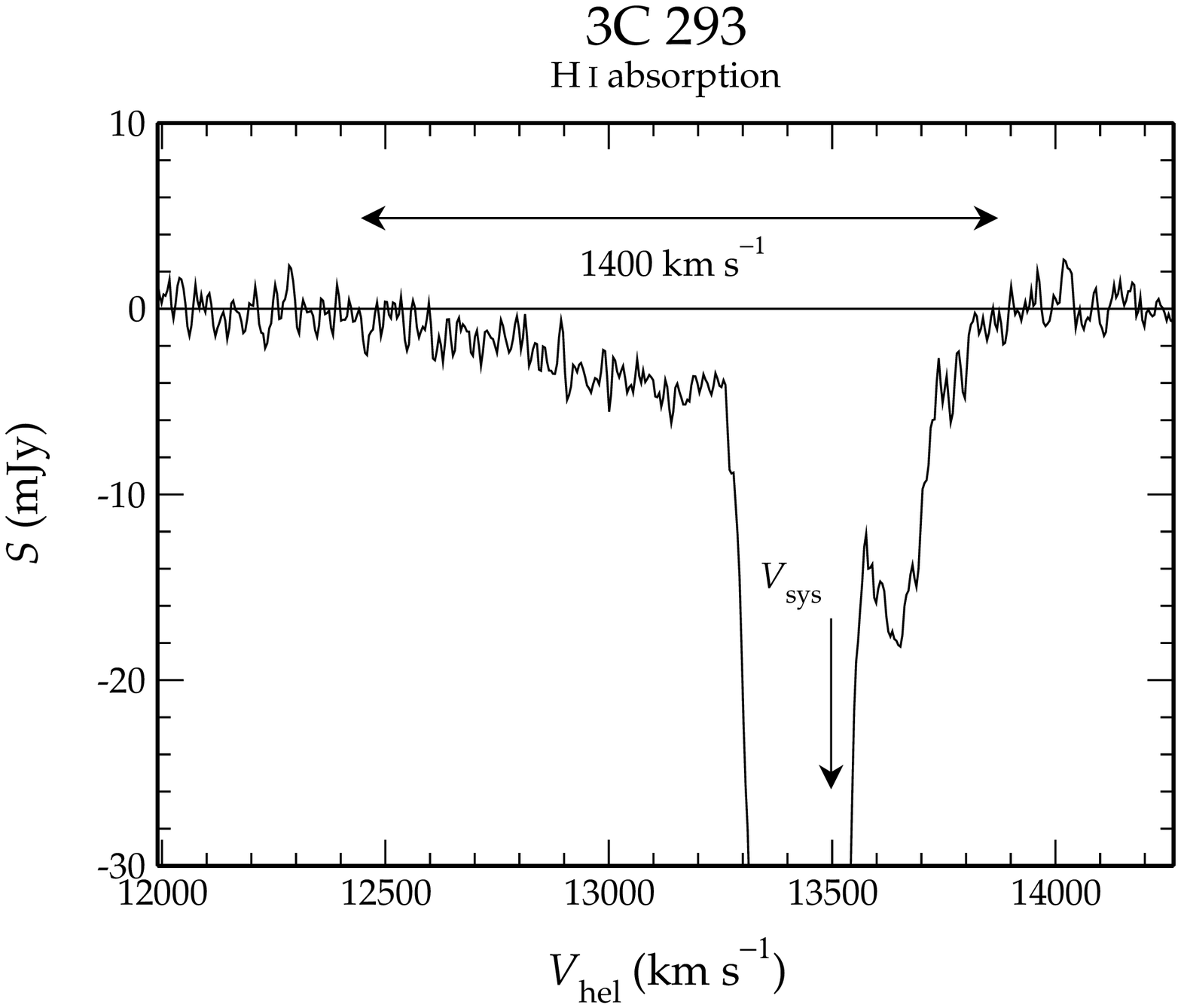,angle=0,width=7cm}
\psfig{figure=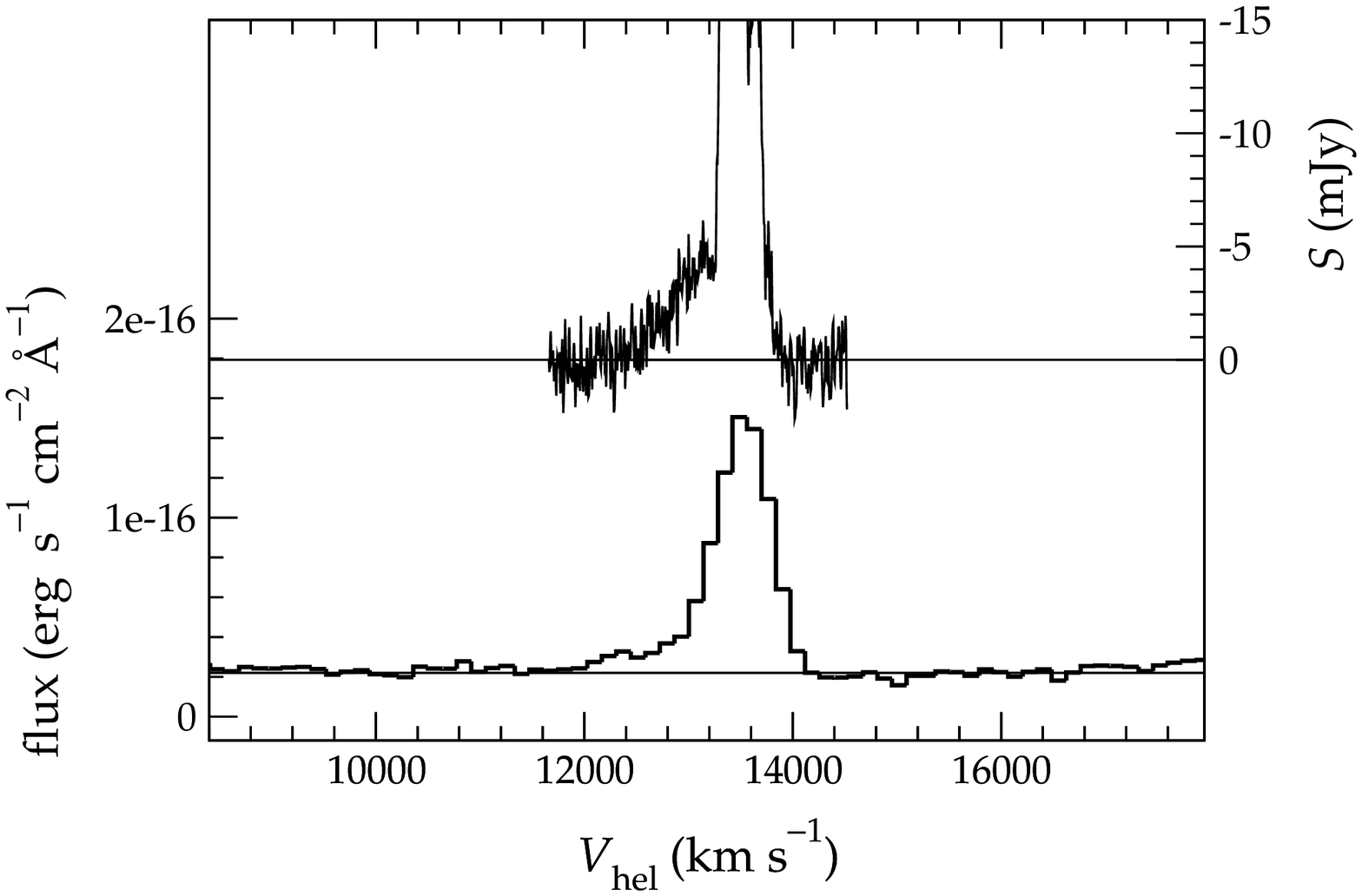,angle=0,width=8cm}}
\caption{
{\sl Left} A zoom-in of the \HI\ absorption spectra of 3C~293 clearly showing
the broad \HI\ absorption. The spectra are plotted in flux (mJy) against
optical heliocentric velocity in \kms.  {\sl Right} Comparison between the
\HI\ absorption (top) and the [OII]3727\AA\ (bottom, from Emonts et al.\ in
prep.) profiles in the radio galaxy 3C~293.  The similarity of the broad,
blueshifted wing in the two profiles is evident.}
\end{figure*}

An even more extreme case of gas outflow has been found in the radio
galaxy 4C~12.50 (PKS~1345+12).  This is a particularly interesting
object as it is a prime candidate for the link between ultraluminous
infrared galaxies (ULIRGs, Sanders \& Mirabel 1996) and radio galaxies
(Evans et al.\  1999). The radio source is confined to a region $<0.1$
arcsec ($\sim 240$ pc) and has all the characteristics of young radio
sources ($<< 10^7$ yr).  The ISM of this radio galaxy is extremely
rich: it is the  brightest far-IR radio galaxy and has a high
molecular gas mass (Evans et al.\ 1999). 
Long slit spectra were taken (using the WHT, see Holt et al.\ 2003 for
details) in order to investigate the impact of the nuclear activity on
the circumnuclear interstellar medium.  The spectra show extended line
emission up to $\sim 20$ kpc from the nucleus, consistent with the
presence of an asymmetric halo of diffuse emission as observed in
optical and infrared images.

At the position of the nucleus, complex emission line profiles are
observed and Gaussian fits to the [OIII] emission lines (see Fig.\ 1
{\sl left}) require three components (narrow, intermediate and broad),
the broadest of which has FWHM$\sim 2000$ \kms\ and is blueshifted by
$\sim 2000$ \kms\ with respect to the halo of the galaxy and the deep
and narrow \HI\ absorption.  This component is interpreted as material
in outflow. A large reddening and high density ($n_{\rm e} >$ 4200
cm$^{-3}$) has also been found for the most kinematically disturbed
component.  4C~12.50, like PKS~1549-79, appears, therefore, to be
a young radio galaxy with nuclear regions that are still enshrouded in
a dense cocoon of gas and dust. The radio jets are now expanding through
this cocoon, sweeping material out of the nuclear regions (Holt et al.\ 2003).

\section{Outflows of neutral hydrogen}

Extremely intriguing is the discovery of a number of radio galaxies
where the presence of fast outflows is associated not only with ionized
but also with {\sl  neutral} gas. This finding gives new and important
insights on the physical conditions of the gaseous medium around an
AGN.

The best example found so far of this phenomenon is the radio galaxy
3C~293 (Morganti et al.\ 2003a). In this galaxy,  very broad \HI\
absorption has been detected against the central regions - this is in
addition to the known and much narrower absorption; Haschick \& Baan
1985, Beswick et al.\ 2002. The absorption profile, obtained using the new
broad band (20~MHz) system available at the Westerbork Synthesis Radio
Telescope, has a full width at zero intensity of about 1400 \kms\ and
most of this broad absorption ($\sim 1000$ \kms) is blueshifted
relative to the systemic velocity. The broad absorption is shallow (the
optical depth is only $\sim 0.15$ \%) and corresponds to a column
density of the \HI\ of $\sim 2 \times 10^{20}\ T_{\rm spin}/100$ K
cm$^{-2}$. This is likely to be a lower limit to the true column
density as the $T_{\rm spin}$ associated with such a fast outflow can
be as large as a few 1000~K (instead of 100~K which is more typical of
the cold, quiescent \HI\ in galaxy disks).  This absorption represents
a fast outflow of {\sl neutral} gas from the central regions of this
AGN.  New optical spectra (see Fig.\ 2b, from Emonts et al.\ in prep.)
show that the optical emission lines also contain a broad component
that is very similar to the broad \HI\ absorption.  This suggests that
the \HI\ and ionized gas outflows may be coming from the same
physical region.

The radio galaxy 4C~12.50 also shows  broad \HI\ absorption (shown in
Fig.\ 1b) when observed using a broad radio band.  The absorption
appears  complex and {\sl extremely broad}.  The full range of
velocities covered by the \HI\ absorption is $\sim 2000$
\kms, {\sl the broadest detected so far in \HI}.  The peak optical
depth of the broad component is only $\tau \sim 0.002$ and the column
density of the full system of shallow \HI\ absorption (assuming a
covering factor is 1) is $\sim 1.7 \times 10^{20} T_{\rm spin}/100K$
cm$^{-2}$.

\begin{figure*}
\centerline{\psfig{figure=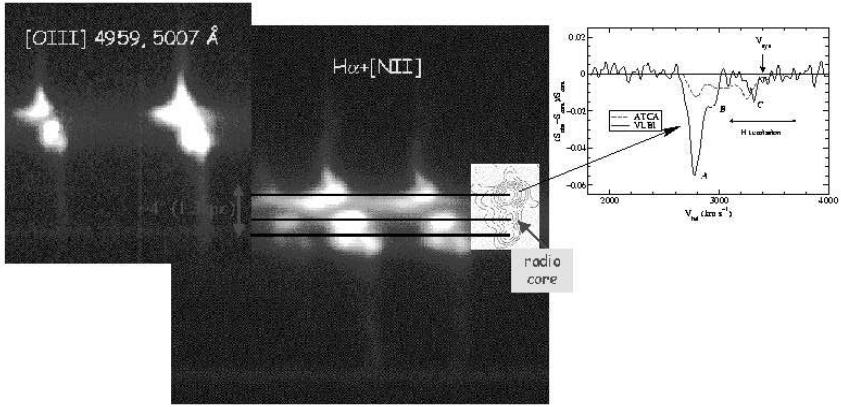,angle=-90,width=14.5cm}}
\caption{
{\sl Left:} a zoom-in of the [OIII] and H$\alpha$+[NII] lines in the Seyfert
IC~5063. The radio image on similar scale is also shown. The highly
blueshifted component close to the bright radio lobe is clearly visible. {\sl
Right:} \HI\ absorption profile profile detected against the NE radio lobe
(from Oosterloo et al.\ 2000).}
\end{figure*}

\subsection{Origin of the \HI\ outflows}

The central question is how {\sl neutral} gas can be associated with
such fast outflows. As discussed in Morganti et al.\ (2003a) for
3C~293, a number of possible hypotheses can be made about the origin
of such outflows. They include, e.g., starburst winds (Veilleux et al. 2002),
radiation pressure from the AGN (Dopita et al.\ 2002), jet-driven
outflow and adiabatically expanding broad emission line clouds (Elvis et
al.\ 2002).

The model that we favour is jet-driven outflow, although
we cannot rule out that one of the other mechanisms is also at work at
some level.  This model assumes that the radio plasma jet hits a
(molecular) cloud in the ISM.  As a consequence of this interaction,
the kinematics of the gas are disturbed by the shocks and the gas is
ionized by it.  Once the shock has passed, part of the gas may have
the chance to recombine and become neutral while it is moving at high
velocities. One problem with this model (considered in detail by
Mellema et al.\ 2002) is that it may not be possible to accelerate the
clouds of gas to the high velocities that we observe, as indicated by
e.g.\ the simulations presented by van Breugel (these proceedings).

To understand which is the more likely mechanism, we now need high
spatial resolution observations to find the exact position of the absorption.
So far this information is available only for the Seyfert galaxy
IC~5063 where it is providing  evidence that the interaction
between the radio jet and a molecular cloud is indeed the cause of the
gas outflow.

\subsection{The case of the Seyfert galaxy IC~5063}

The southern Seyfert galaxy IC~5063 was the first AGN where a fast
outflow of neutral hydrogen was detected (Morganti et al.\ 1998). VLBI
observations have confirmed that the broad absorption is detected
against the bright radio lobe (i.e.\ {\sl not} against the nucleus)
situated at $\sim 1.3$ kpc from the nucleus, as shown in Fig.\ 3a (see
also Oosterloo et al.\ 2000).

Deep optical spectra have been taken with the ESO-NTT to compare the
kinematics of the ionized gas with that of the neutral hydrogen component in
order to study the mechanism that could drive this outflow. Preliminary
results are presented in Morganti et al. (2003b). The data reveal extremely
complex gas kinematics as can be seen in Fig.\ 3b. This includes the presence
of an outflow of ionised gas (with velocities of several hundred km/s) at the
location of the brighter radio lobe.  The amplitude of this outflow is
strikingly similar to that of the \HI.  Because of its location, we consider
the interaction between the radio jet and the ISM to be the most likely
mechanism for the extreme kinematics.

\section{Conclusions}

Our observations show that gas outflows of both ionized gas and neutral
hydrogen are found in some radio galaxies.  So far the indications are that
these outflows are present in objects that are either in the early-stage of
their evolution (like 4C~12.50) or, perhaps, in a phase of re-started activity
(as we think it is the case for 3C~293). Moreover, another characteristic that
appears to be common among these objects is the presence of a young stellar
population component (Tadhunter et al.\ 2003). \\ Although it is difficult to
attribute the outflow to a starburst wind (as the young stellar population is
usually relatively old, 0.5-2 Gyr), the presence of such a young stellar
component could be an indication that the galaxy is indeed in a particular
stage of its evolution, where large amounts of gas/dust - likely from the
merger that triggered the activity - are still present in the inner region and
the radio jet is strongly interacting with it. \\ A more systematic search for
fast gas outflows in radio galaxies is now in progress.

\end{document}